\documentclass{acm_proc_article-sp}
\usepackage{graphicx}
\usepackage{tabularx}
\newcolumntype{Y}{>{\centering\arraybackslash}X}
\begin{document}

\conferenceinfo{}{Bloomberg Data for Good Exchange 2017, NY, USA}

\title{Predictors of Re-admission for Homeless Families in New York City: The Case of the Win Shelter Network}

\numberofauthors{8}
\author{
\alignauthor
Constantine Kontokosta\\
\affaddr{Asst. Prof. (Urban Informatics)\\New York University\\(CUSP \& Tandon)}\\
\email{ckontokosta@nyu.edu}
\alignauthor
Boyeong Hong\\
\affaddr{Civic Analytics Fellow\\New York University\\(CUSP)}\\
\email{boyeong.hong@nyu.edu}
\alignauthor
Awais Malik\\
\affaddr{PhD Candidate (Civil Eng.)\\New York University\\(CUSP \& Tandon)}\\
\email{awais.malik@nyu.edu}
\and
\alignauthor
Ira M. Bellach\\
\affaddr{Associate Vice President, IT\\Women in Need NYC}
\email{ibellach@winnyc.org}
\alignauthor
Xueqi Huang\\
\affaddr{New York University\\(CUSP)}\\
\email{xh895@nyu.edu}
\alignauthor
Kristi Korsberg\\
\affaddr{New York University\\(CUSP)}\\
\email{kk3374@nyu.edu}
\and
\alignauthor
Dara Perl\\
\affaddr{New York University\\(CUSP)}\\
\email{dp1618@nyu.edu}
\alignauthor
Avikal Somvanshi\\
\affaddr{New York University\\(CUSP)}\\
\email{as10724@nyu.edu}
}

\maketitle
\begin{abstract}
New York City faces the challenge of an ever-increasing homeless population with almost 60,000 people currently living in city shelters. In 2015, approximately 25\% of families stayed longer than 9 months in a shelter, and 17\% of families with children that exited a homeless shelter returned to the shelter system within 30 days of leaving. This suggests that "long-term" shelter residents and those that re-enter shelters contribute significantly to the rise of the homeless population living in city shelters and indicate systemic challenges to finding adequate permanent housing. Women in Need (Win) is a non-profit agency that provides shelter to almost 10,000 homeless women and children (10\% of all homeless families of NYC), and is the largest homeless shelter provider in the City. This paper focuses on our preliminary work with Win to understand the factors that affect the rate of readmission of homeless families at Win shelters, and to predict the likelihood of re-entry into the shelter system on exit. These insights will enable improved service delivery and operational efficiencies at these shelters. This paper describes our recent efforts to integrate Win datasets with city records to create a unified, comprehensive database of the homeless population being served by Win shelters. A preliminary classification model is developed to predict the odds of readmission and length of shelter stay based on the demographic and socioeconomic characteristics of the homeless population served by Win. This work is intended to form the basis for establishing a network of ``smart shelters" through the use of data science and data technologies.
\end{abstract}



\section{Introduction}
New York City faces the challenge of an ever-increasing homeless population with almost 60,000 people currently living in city shelters \cite{DHScensus2017}. A New York City Department of Investigation (DOI) probe of Department of Homeless Services (DHS) shelters in 2015 found ``serious deficiencies'' in the shelters for families with children, citing unsafe and unhealthy living conditions, and recommending aggressive and immediate reforms \cite{peters2015}. Consequently, the city plans to spend \$2.3 billion on efforts to improve the quality of life of the homeless population in the fiscal year 2017 \cite{deBlasio2017}. In 2015, 17\% of families with children that exited a homeless shelter returned to the shelter system within 30 days of leaving \cite{ICPH2017}. This highlights the problem that the sharp increase in the homeless population residing in the shelter system is exacerbated by long-term residents and those that re-enter shelters, as much as (if not more than) new homeless families \cite{ICPH2017}.

Women in Need (Win) is a non-profit agency that provides shelter to almost 10,000 homeless women and children (10\% of all homeless families in NYC), and is the city's largest homeless shelter provider \cite{WinNYC}. Win's goal is to enable homeless women to become self-sufficient as they transition from a Win shelter into permanent housing. In order to deliver better services and reduce the overall homeless population, Win has provided us with anonymized sociodemographic data for its residents for the past five years. Our goal in this paper is to understand the factors that significantly affect the rate of readmission into Win shelters after initial exit, a key positive indicator.

This paper begins with a brief review of previous studies focused on the state of homelessness in New York City and predictors of homeless family outcomes. The next section describes the data provided by Win, and our recent efforts to integrate these datasets to create a unified database of the homeless population served by Win shelters. We proceed to develop a predictive model of homeless family outcomes, starting with the odds of readmission back into the shelter system. Considering the nature of the data, we discuss some of salient ethical and privacy concerns associated with working with such data, and how we will address these concerns. Since this is a work in progress, we discuss our next steps and future research, including predicting length of stay for each homeless family. This paper concludes with our vision of creating a network of ``smart shelters'' in New York City, and offers suggestions for how our work can serve as a starting point for implementing such a network.

\section{Background}
Homelessness in New York City has reached its highest levels since the Great Depression, with more than 127,000 people sleeping in the NYC shelter system at some point during the fiscal year 2016 \cite{CoalitionforHomeless}. The New York City Department of Homeless Services (DHS) provides temporary emergency shelter to the homeless, and assists in their transition into permanent housing. One of DHS's main partners is Women-in-Need, Inc. (Win) - the City's largest homeless shelter provider, serving more than 10,000 homeless people (10\% of all homeless families in NYC), including more than 800 families and 6,000 children \cite{WinNYC}. According to the data provided by Win, their shelter network houses close to 4,700 people every night. The data also show that in March 2017 46\% of Win's clients were employed, making approximately \$1,420 per month, only slightly more than the federal poverty line.

A few studies have examined the factors associated with homeless families returning to the shelter system. Wong et al. \cite{wong1997} researched readmission patterns of New York City's homeless population, and analyzed certain demographic characteristics, time variables, and reasons for homelessness with respect to the likelihood of readmission. The authors used eight years of client data from the Homeless Emergency Referral System (HOMES) database, and found that five demographic features (age, family size, race and ethnicity, pregnancy status, and receiving public assistance) were the most significant predictors of shelter readmission \cite{wong1997}. Additionally, the study found that families that exited into subsidized housing stayed in the shelter system for a longer time, but had a lower rate of readmission.

Shinn et al. \cite{shinn1998} examined predictors of entry into homeless shelters and subsequent housing stability. The authors surveyed 266 New York City families as they requested shelter, and compared them with a random sample of 298 families from the welfare caseload \cite{shinn1998}. The responding families were reinterviewed five years later to evaluate their housing stability, and families with prior history of shelter use were removed from the follow-up study \cite{shinn1998}. The authors found that demographic characteristics and housing conditions were the most significant risk factors affecting shelter entry, with `enduring poverty' and `disruptive social experiences' also important conditions. Access to subsidized housing was the most crucial factor in long-term housing stability: the odds of housing stability were 20 times greater for recipients of housing subsidies or tenants in subsidized housing \cite{shinn1998}.

Recently, the Institute for Children, Poverty and Homelessness \cite{ICPH2017} provided an in-depth analysis of demographic patterns and social dynamics of family homelessness in New York City. The institute suggests that the growing population of the shelter system is primarily caused by long-term staying clients and repeated entries, not by new homeless families \cite{ICPH2017}. The is supported by the fact that 17\% of families with children that exited a homeless shelter returned to the shelter system within 30 days of leaving \cite{ICPH2017}. This report suggests that in order to reduce homelessness in New York City, the families returning immediately to homeless shelters need to be provided more stable housing solutions upon exiting shelters \cite{ICPH2017}, such as receiving subsidized housing \cite{shinn1998}. This study provides an important foundation for our analysis.

\section{Data \& Methodology}
Win provided anonymized social and demographic data on the homeless population that entered and exited its shelters over the past five years (FY 2013 - FY 2017). The data were split into three different files: the first dataset included information about client demographics (Demographic Data) such as gender, race and ethnicity, education level, employment status, income etc.; the second identified reasons for client exit (Exits Data); and the third summarized incidents involving clients that occurred while in the shelter (Incidents Data).

A significant challenge in merging these datasets was the presence of three separate identifiers representing the individual (CARES ID), the individual's family (FAMILY ID), and the individual's current case number (CASE ID). We created a new identifier for each unique individual ('ID combo') concatenating the three identifiers associated with each record. It is worth noting that potential Win clients undergo a 14 day case review to determine eligibility when they apply to the shelter network. Those cases that were not admitted to the Win system were removed from the unified dataset for the purpose of this analysis.

The total length of stay for each individual was calculated by summing the lengths of stay for each distinct period of residence for the same individual. An individual with more than one distinct period of residence was considered a multi-entry client. The integrated dataset featured 6,779 unique individuals with 1,289 (19\%) classified as multi-entry clients, i.e. they were readmitted to the shelter more than once in the last five years. Eviction, discord, domestic violence and overcrowding were the top four reasons that Win clients needed shelter, while 48-hour curfew violations, family reunification, and independent living were the predominant reasons for Win clients exits.

Table 1 lists the predictors extracted from the unified dataset for our preliminary estimation of the odds of reentry for Win clients. The income variable was excluded from our models because nearly half of the individuals had missing values for this field, and thus we were unable to determine if the client had no income or the value was missing. The target variable is coded as a binary (readmit = 0 for single entry client, and readmit = 1 for multi-entry). 81\% of the individuals in the integrated dataset were single-entry clients (readmit = 0). A logistic regression model was used to establish a benchmark for the predictive power of selected features. Subsequently, a Gradient Boosting Classifier was applied using various ratios for minority oversampling to improve model performance. 

\begin{table}[h!]
\caption{Description of Extracted Predictors}
\smallskip
\def\arraystretch{1.25}
\begin{center}
\begin{tabularx}{\linewidth}{ |X|X| }
 \hline
 \textbf{Predictor} & \textbf{Description}\\
 \hline
 Age & Continuous Variable\\
 \hline
 Income & {Continuous Variable \newline Nearly 50\% of data missing}\\
 \hline
 Race & Categoric Variable\\ 
 ~ & 0: White\\ 
 ~ & 1: Black\\ 
 ~ & 2: Hispanic\\ 
 ~ & 3: Other\\ 
 \hline
 Family Type & Categoric Variable\\
 ~ & 0: Single\\
 ~ & 1: Adult Families\\
 ~ & 2: Families with Children\\
 \hline
 Reason for Homelessness & Categoric Variable \\
 ~ & 0: Eviction\\
 ~ & 1: Discord\\
 ~ & 2: Domestic Violence\\
 ~ & 3: Overcrowding\\
 ~ & 4: Other\\
 \hline
 Employment Status & Categoric Variable\\
 ~ & 0: Unemployed\\
 ~ & 1: Employed\\
 ~ & 2: Unknown\\
 \hline
 Citizenship Status & Categoric Variable \\
 ~ & 0: Unknown\\
 ~ & 1: Citizen\\
 ~ & 2: Non-Resident\\
 ~ & 3: Undocumented\\
 \hline
\end{tabularx}
\end{center}
\end{table}

\section{Preliminary Results}
Figure 1 shows the ROC curve for the logistic regression model. The AUC is very low (0.57), which suggests the need for a non-parametric model to deal with non-linear relationships. Therefore, a Gradient Boosting Classifier is applied to the dataset. Since 81\% of the dataset is single entry clients, we need a sample-balancing mechanism, so we apply the Synthetic Minority Over-sampling Technique (SMOTE) \cite{chawla2002} with various SMOTE ratios ranging from the original distribution of the data to 1.0.

\begin{figure}[h!]
  \includegraphics[width=\linewidth]{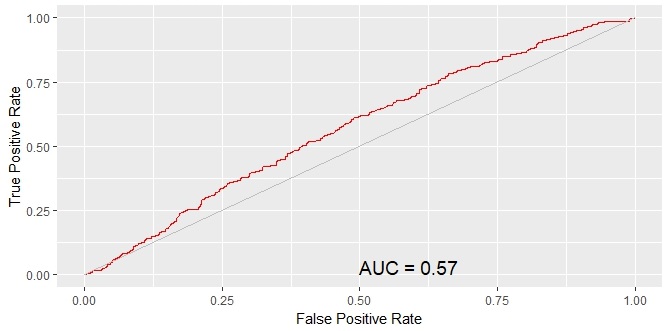}
  \caption{ROC Curve for Logistic Regression Model}
\end{figure}

Figure 2 shows the ROC curve for the Gradient Boosting Classifier with a SMOTE ratio of 1.0. This is the best performing model (see Table 2) because it has the highest recall (sensitivity). We use recall as our model performance evaluation metric because our priority is to reduce the number of False Negatives (homeless individuals classified as single entry by the model, but who actually return to the shelter after their first exit). An AUC of 0.72 is a reasonable starting point, and the addition of more suitable indicators (such as the length of stay during first visit) may further enhance the model performance.

\begin{figure}[h!]
  \includegraphics[width=\linewidth]{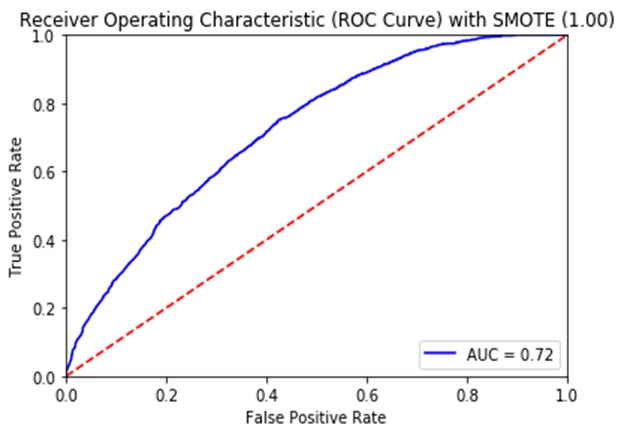}
  \caption{ROC Curve for Gradient Boosting Classifier with SMOTE Ratio of 1.0}
\end{figure}

\begin{table*}
\centering
\caption{Performance Evaluation Metrics of Gradient Boosting Classifier using Various SMOTE Ratios}
\smallskip
\def\arraystretch{1.5}
\begin{tabularx}{\textwidth}{|l|XXXXXXXXX|}
\hline
SMOTE Ratio & Original & 0.3 & 0.4 & 0.5 & 0.6 & 0.7 & 0.8 & 0.9 & 1.0 \\
\hline
Accuracy & 0.83 & 0.81 & 0.79 & 0.78 & 0.78 & 0.78 & 0.78 & 0.79 & 0.79 \\
True Positives & 128 & 83 & 127 & 171 & 230 & 357 & 448 & 549 & 588 \\
False Negatives & 1161 & 1206 & 1162 & 1118 & 1059 & 932 & 841 & 740 & 701 \\
False Positives & 14 & 6 & 59 & 102 & 223 & 496 & 700 & 958 & 1037 \\
True Negatives & 5476 & 5484 & 5431 & 5388 & 5267 & 4994 & 4790 & 4532 & 4453 \\
AUC & 0.79 & 0.76 & 0.75 & 0.74 & 0.73 & 0.73 & 0.73 & 0.73 & 0.72 \\
Sensitivity & 0.099 & 0.064 & 0.099 & 0.133 & 0.178 & 0.277 & 0.348 & 0.426 & 0.456 \\
\hline
\end{tabularx}
\end{table*}

\section{Conclusions \& Next Steps}
We have presented our preliminary findings for predicting readmission of homeless families back into Win shelters in New York City. Predictors include age, race, family type, employment status, citizenship status, and reason for homelessness. Our model of choice is a Gradient Boosting Classifier with SMOTE ratio of 1.0, which gives an AUC of 0.72 and recall of 0.456. Model performance is evaluated using recall, since we want to reduce the number of false negatives, or at least have the number of false negatives relatively lower than the number of false positives.

Further improvements in the model include accounting for the interactions between continuous variables (such as age) and the categorical predictors, as well as extracting more relevant indicators from the Win datasets (such as length of stay during first period of residence and whether the client experienced a traumatic incident during their initial residency). We also want to predict the length of stay for both single entry clients and multi-entry clients independently. Current attempts at predicting length of stay for the unified dataset using linear regression methods did not yield insightful results, and were, therefore, not presented in this paper.

Finally, we believe our work with Win will serve as a benchmark for establishing data-driven operations within the largest homeless shelter provider in New York City. Furthermore, our currnet work with Win is foundation for a long-term partnership to form a network of ``smart shelters'' across the City. This network will serve to support improved outcomes for homeless families both during and after their time in the shelter system by appropriately using data and technology to address persistent challenges.

\section{Acknowledgments}
Our thanks to Win for their partnership in this research, especially to Ira M. Bellach and Meghan Linehan for their assistance and guidance. This work is supported, in part, by the John D. and Catherine T. MacArthur Foundation. This research has been conducted in accordance with NYU Institutional Review Board approval IRB-FY2017-1135. Any opinions, findings, and conclusions expressed in this paper are those of the authors and do not necessarily reflect the views of any supporting institution. All errors remain the responsibility of the authors.

\nocite{*}
\bibliographystyle{abbrv}
\bibliography{references}

\end{document}